\documentclass[conference]{IEEEtran}
\IEEEoverridecommandlockouts

\usepackage{graphicx}
\usepackage{amsmath,amsfonts,amssymb}
\usepackage{url}
\usepackage{cite}
\usepackage{xurl}
\usepackage[hidelinks]{hyperref}
\usepackage{listings}
\usepackage{color}
\usepackage{array}
\usepackage{booktabs}
\usepackage{multirow}
\usepackage{tikz}
\usepackage{comment}
\usepackage{balance}

\usetikzlibrary{arrows.meta,positioning}

\title{CyTReX: Explainable AI-Based Cybersecurity Threat Reasoning Framework for DER Networks}

\author{\IEEEauthorblockN{Damilola Popoola, Souradeep Bhattacharya, and Manimaran Govindarasu}
\IEEEauthorblockA{\textit{Department of Electrical and Computer Engineering} \\
\textit{Iowa State University, Ames, IA 50011}\\
Email: popson@iastate.edu, sbhatta@iastate.edu, gmani@iastate.edu}
}

\begin{document}

\maketitle

% ----------------------------------------------------------------
% ABSTRACT
% Acronyms defined here: AI, MCP, CTI, LLM, DER, ADS, SOC, SHAP, CyTReX
% ----------------------------------------------------------------
\begin{abstract}
Distributed Energy Resource (DER) environments rely on network communication protocols to coordinate control commands, measurements, and device states across edge assets and cloud systems. Edge anomaly detection systems (ADS) monitor this traffic to identify deviations from normal communication behavior, flagging suspicious flows for further investigation. When the ADS flags abnormal network traffic, a single attack label is often insufficient for operational response: the label reports the detector's selected class but does not expose alternative threat interpretations that may warrant investigation. This paper presents Cybersecurity Threat Reasoning with Explainable Artificial Intelligence (CyTReX), an evidence-grounded threat reasoning framework for DER security that transforms network-level anomaly alerts into ranked, analyst-facing threat hypotheses designed to support Security Operations Center (SOC) triage and investigation. CyTReX constrains large language model (LLM) reasoning through a structured evidence packet, defined as a consolidated record of detection outputs, model explanations, and cyber threat intelligence (CTI) context. The evidence packet integrates edge-layer anomaly detection evidence, cloud reasoning layer attack interpretation, Shapley Additive Explanations (SHAP) network-feature attributions, surrogate decision rules, and Model Context Protocol (MCP)-enabled CTI enrichment. This ensures that every ranked hypothesis and attack-tree branch is traceable to explicit evidence rather than free-form LLM inference, and that incomplete or conflicting evidence is communicated rather than suppressed. Evaluation across five configurations shows that additional reasoning components improve hypothesis specificity, evidence traceability, and analytical grounding, with the complete pipeline providing the richest evidence-grounded reasoning context.
\end{abstract}


% ----------------------------------------------------------------
% SECTION I — INTRODUCTION
% The main text independently defines its acronyms because the abstract is standalone.
% ----------------------------------------------------------------
\section{Introduction}
\bstctlcite{IEEEexample:BSTcontrol}

Distributed Energy Resources (DER) environments rely on network communication protocols such as Distributed Network Protocol 3 (DNP3) to coordinate control commands, device states, and measurements across edge assets, aggregators, and cloud systems \cite{zografopoulos2023der}. Abnormal DNP3 traffic may reflect replay attacks, unauthorized control commands, or denial-of-service (DoS) behavior, and multi-class identification requires distinguishing among several attack types from network-flow features \cite{dangwal2024effective,abdelkhalek2022mlbased}. Critically, a single-class detection label reports only the selected class and does not communicate which alternative threat interpretations remain plausible or what additional evidence would distinguish them.

Detectors based on artificial intelligence (AI) and machine learning (ML) improve coverage but do not resolve this gap \cite{ahmad2021network,hermosilla2025explainable}. Without understanding which network features drove the alert, how they relate to known adversary techniques, and which alternative behaviors warrant investigation, a Security Operations Center (SOC) analyst has no reliable basis for triage, escalation, or response.

This paper presents CyTReX, a threat reasoning framework for DER security that converts ML-generated network anomaly alerts into ranked, evidence-supported threat hypotheses. CyTReX constrains large language model (LLM) reasoning through a structured evidence packet that integrates edge anomaly evidence, cloud reasoning layer attack interpretation, Shapley Additive Explanations (SHAP) network-feature attributions, surrogate decision rules, and Model Context Protocol (MCP)-enabled cyber threat intelligence (CTI) enrichment. This ensures that every ranked hypothesis and attack-tree branch is traceable to explicit evidence rather than free-form LLM inference, and that incomplete or conflicting evidence is communicated rather than suppressed. The main contributions are: \textbf{\textit{(i)}} an edge-cloud reasoning workflow that transforms DER network anomalies into ranked threat hypotheses to support SOC analyst triage; \textbf{\textit{(ii)}} a unified evidence packet consolidating network-feature detection outputs, model explanations, CTI context, and uncertainty as structured LLM input; and \textbf{\textit{(iii)}} a five-configuration DNP3 case study evaluating each reasoning component's contribution to hypothesis specificity, evidence traceability, and unsupported-claim reduction.

% ----------------------------------------------------------------
% SECTION II — BACKGROUND AND MOTIVATION
% DER, DNP3, CTI, LLM, MCP already defined — do NOT redefine
% ----------------------------------------------------------------
\section{Background and Motivation}

\subsection{DER-Integrated Grid Architecture}

\begin{figure}[t]
    \centering
    \includegraphics[width=\columnwidth]{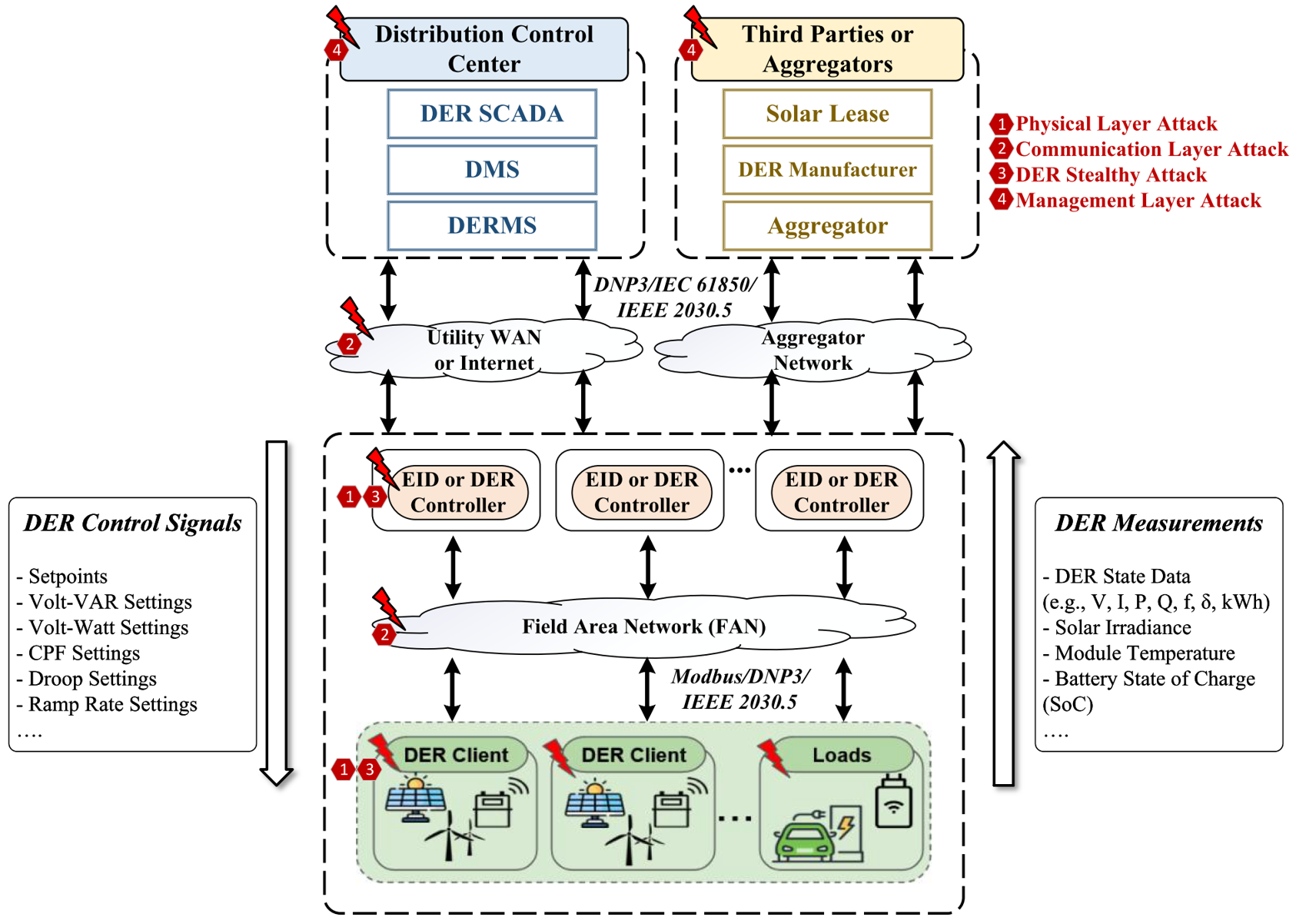}
    \vspace{-20pt}
    \caption{Representative DER-integrated grid communication architecture and attack surfaces \cite{11224995}.}
    \label{fig:der-architecture}
    \vspace{-15pt}
\end{figure}

Modern grid environments increasingly rely on DER assets, edge devices, communication gateways, control applications, and cloud-based analytics for monitoring and operation \cite{zografopoulos2023der}. DER resources such as solar photovoltaic systems, battery storage units, electric vehicles, and smart inverters exchange measurements, commands, and status information using standardized communication protocols, including DNP3 \cite{ieee1815standard}. As shown in Fig.~\ref{fig:der-architecture}, DER-integrated architectures connect DER clients, edge intelligent devices (EIDs), field-area networks, aggregators, and distribution control systems through multiple communication paths \cite{epri_2019}.

Because DER communication tightly couples network traffic, control commands, and device state, the same network-level anomaly can require different interpretations depending on protocol, device, and operational context. A single-class detector reports its selected class but does not enumerate alternative hypotheses or the evidence needed to distinguish them. Edge devices observe communication close to the asset and forward alerts to cloud-side analytics frameworks for aggregation and interpretation, but neither layer alone supplies the full evidence context needed for such reasoning \cite{sugunaraj2026derms}. This motivates the need for SOC-oriented reasoning frameworks that go beyond the flagged label to examine what else the traffic may indicate.

\subsection{DER Cybersecurity Context and Threat Reasoning}

DER cybersecurity protects the communication, computation, and control processes that support reliable operation of distributed energy assets \cite{sridhar2011cyber, ashok2017cyber}. In DER-integrated environments, security events can affect control actions, telemetry visibility, device coordination, and operator decision-making \cite{sen2025cybersecurity}. As these environments become more digitalized, field devices and edge monitoring systems generate network traffic data that may reveal abnormal communication patterns. However, a single detection label does not show which alternative threat interpretations remain plausible or how the affected device and command context should influence triage.

Cybersecurity threat reasoning addresses this interpretation gap by moving beyond the detection label toward structured explanation. Rather than treating an alert class as the final output, threat reasoning connects network communication behavior, model interpretation, and contextual knowledge to determine what the alert may indicate and what alternative behaviors may remain plausible \cite{sen2024towards}. This is important in DER environments because the same abnormal traffic pattern may reflect different operational consequences depending on the affected device, command type, and available supporting evidence.

\subsection{Need for Evidence-Grounded Reasoning}

Evidence-grounded reasoning is important for DER cybersecurity because operational decisions cannot rely only on anomaly scores or isolated class labels \cite{zibi2025systematic}. Without supporting evidence, a SOC analyst cannot determine whether an alert reflects a confirmed threat, a benign operational condition, or an uncertain behavior that requires further investigation, and cannot assess which secondary threat hypotheses merit further investigation.

By linking each threat interpretation to observable network behavior, model explanations, CTI context, and evidence gaps, evidence-grounded reasoning improves the transparency and usefulness of DER security alerts \cite{hermosilla2025explainable}. It helps analysts understand why a hypothesis is plausible, which alternative explanations remain possible, and what additional evidence is needed before response actions are taken. This paper adopts this evidence-grounded view by constraining LLM reasoning through a structured evidence packet, ensuring that every ranked hypothesis is traceable to explicit evidence and that incomplete or conflicting evidence is communicated rather than suppressed.

\begin{table*}[t]
\renewcommand{\arraystretch}{1.3}
\caption{Comparison of Cybersecurity Threat Reasoning and Analysis Frameworks}
\vspace{-0.2cm}
\centering
\label{tab:related_works}
\resizebox{\textwidth}{!}{
\begin{tabular}{c c c c c c c}

\hline\hline \\[-4mm]
\multicolumn{1}{c}{\textbf{Framework / Approach}} & \multicolumn{1}{c}{\textbf{ML Anomaly Detection}} & \multicolumn{1}{c}{\textbf{Explainable AI (XAI)}} & \multicolumn{1}{c}{\textbf{CTI Enrichment}} & \multicolumn{1}{c}{\textbf{LLM-Assisted Reasoning}} & \multicolumn{1}{c}{\textbf{Ranked Hypotheses}} & \multicolumn{1}{c}{\textbf{Grid/Industrial Control Systems (ICS) Focus}} \\ \hline

ML-Based Anomaly Detection System (ML-ADS) \cite{abdelkhalek2022mlbased} & $\checkmark$ & $\times$ & $\times$ & $\times$ & $\times$ & $\checkmark$ \\

CySIDER \cite{cysider2025} & $\checkmark$ & $\times$ & $\checkmark$ & $\times$ & $\times$ & $\checkmark$ \\

XAI-Forensic Intrusion Detection System (IDS) \cite{hermosilla2025explainable} & $\checkmark$ & $\checkmark$ & $\times$ & $\times$ & $\times$ & $\times$ \\

ExCyDER \cite{excyder2026} & $\checkmark$ & $\checkmark$ & $\times$ & $\times$ & $\times$ & $\checkmark$ \\

LLM Cybersecurity Survey \cite{ferrag2025generative} & $\times$ & $\times$ & $\checkmark$ & $\checkmark$ & $\times$ & $\times$ \\

SEvenLLM \cite{ji2024sevenllm} & $\times$ & $\times$ & $\checkmark$ & $\checkmark$ & $\times$ & $\times$ \\

\textbf{CyTReX} & $\checkmark$ & $\checkmark$ & $\checkmark$ & $\checkmark$ & $\checkmark$ & $\checkmark$ \\
              
\hline\hline
\end{tabular}
}
\vspace{-0.4cm}
\end{table*}

% ----------------------------------------------------------------
% SECTION III — RELATED WORK
% ML, SHAP, LIME, CTI, LLM already defined — do NOT redefine
% Split into four focused paragraphs; added SHAP citation
% ----------------------------------------------------------------
\section{Related Work}

Machine learning-based anomaly detection has been widely studied for networked cyber-physical systems, industrial control systems, and grid communication environments \cite{ahmad2021network,elsaid2024optimized,roopraj2025intrusion}. In DNP3-enabled DER networks, supervised classifiers have also been used to distinguish multiple cyber/physical attack classes \cite{abdelkhalek2022mlbased}. However, anomaly detection mainly identifies abnormal behavior and does not provide sufficient attack semantics. Closed-set classifier accuracy does not by itself establish reliability for attack behavior excluded from training; unknown-attack performance must be evaluated separately \cite{ozdogan2025structured}. These limitations have motivated DER cybersecurity research that moves beyond detection toward situational intelligence and higher-level alert interpretation \cite{cysider2025}. While such work improves the operational context of DER security monitoring, there remains a need for threat reasoning methods that explicitly connect ML evidence to CTI context.

Explainable AI methods, including SHAP \cite{lundberg2017unified} and Local Interpretable Model-agnostic Explanations (LIME), have been used to interpret anomaly detection decisions and identify influential features \cite{hermosilla2025explainable,muhammad2025lime}. However, feature explanations alone do not show whether the evidence is consistent with a plausible cyber threat hypothesis or how it should be combined with operational context, rule evidence, and threat intelligence. Recent DER-focused work has addressed part of this issue through self-verifying anomaly detection, where explanation consistency is used to assess the reliability of anomaly evidence \cite{excyder2026}. These efforts show the importance of moving from feature-level explanation toward evidence-aware interpretation in grid cybersecurity.

CTI repositories provide external context for interpreting observed cyber behavior. Structured Threat Information Expression (STIX) and Trusted Automated Exchange of Intelligence Information (TAXII) support standardized CTI representation and exchange \cite{oasis_stix,oasis_taxii}. MITRE Adversarial Tactics, Techniques, and Common Knowledge (ATT\&CK) for ICS and Common Attack Pattern Enumeration and Classification (CAPEC) provide complementary behavior-level knowledge about adversary tactics, techniques, and procedures (TTPs), attack patterns, and potential mitigations \cite{mitre_attack_data,capec}. While these sources can enrich alert interpretation, CTI must be grounded in observed evidence to avoid unsupported associations, especially when direct packet-level or operational evidence is incomplete.

LLMs and generative AI have recently been explored for cybersecurity decision support, incident summarization, threat analysis, CTI synthesis, vulnerability interpretation, and analyst-assistance workflows \cite{ferrag2025generative,ren2025large,ji2024sevenllm,srinivas2025llmsoc}. These models are useful because they can organize heterogeneous security information, summarize complex evidence, and generate human-readable explanations that support investigation. In cyber-physical and grid-security settings, such capabilities are relevant because alerts may involve protocol behavior, device context, operational impact, and external threat knowledge \cite{yigit2025genai}. However, LLM outputs can be sensitive to prompt design and may overgeneralize, introduce unsupported CTI associations, or present uncertain interpretations as confirmed conclusions when evidence constraints are weak \cite{ali2025beyond,ferrag2025generative}.

Table~\ref{tab:related_works} summarizes the landscape of related works across these domains. Taken together,  existing methods either focus on isolated detection, provide feature-level explanations without operational context, or apply LLMs without strict evidence grounding. This leaves a critical gap for alert interpretation methods that connect low-level cyber-physical evidence with ranked, evidence-grounded threat hypotheses. 
%CyTReX addresses that gap by using the LLM exclusively as a reasoning and structuring layer, operating only after detection evidence, explicit model explanations, and CTI context have been assembled into a unified evidence packet.

% ----------------------------------------------------------------
% SECTION IV — PROPOSED METHODOLOGY
% CTI, LLM, MCP already defined — do NOT redefine
% ----------------------------------------------------------------
\section{Proposed Methodology}

\subsection{Framework Overview}

\begin{figure}[t]
    \centering
    \includegraphics[width=\columnwidth]{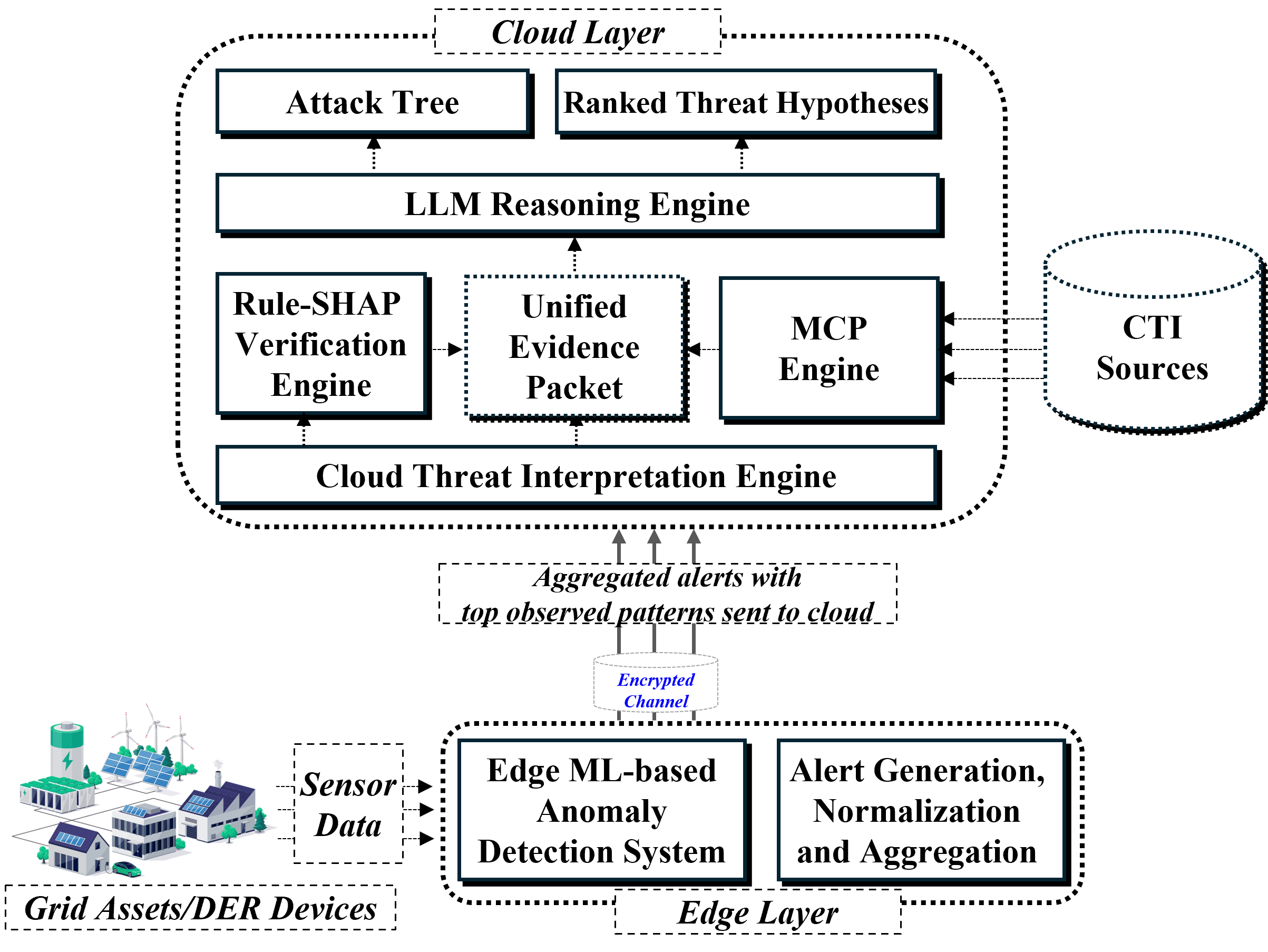}
    \vspace{-20pt}
    \caption{CyTReX framework showing transmission of edge-generated alert evidence to the cloud for evidence-grounded threat reasoning and hypothesis generation}
    \label{fig:framework}
    \vspace{-15pt}
\end{figure}

The proposed CyTReX framework provides an explainable and evidence-grounded cyber threat reasoning workflow for DER-integrated grid security. As shown in Fig.~\ref{fig:framework}, CyTReX transforms cyber-physical alerts from communication or telemetry behavior into structured attack trees with ranked threat hypotheses. Rather than treating detection outputs as final decisions, the proposed framework organizes alert evidence through an edge-cloud pipeline, where the edge layer generates early abnormality evidence and the cloud reasoning layer performs threat interpretation, model explanation, CTI enrichment, and LLM-based attack-tree reasoning to support uncertainty-aware analyst investigation.

Operationally, CyTReX informs SOC triage by converting a flagged DER network alert into a structured investigation object. The edge layer first identifies abnormal traffic and forwards the alert evidence to the cloud reasoning layer. The cloud layer then adds attack-class interpretation, model explanation evidence, CTI context, and evidence-quality signals before the LLM structures the information into ranked hypotheses and attack-tree branches. This output helps analysts see the most supported threat interpretation, related secondary possibilities, relevant CTI and mitigation context, and remaining evidence gaps before deciding whether to escalate, investigate further, or approve a response action.

\subsection{Edge Evidence Generation Layer}

The edge layer provides the first level of evidence for the reasoning workflow. It observes communication or telemetry behavior close to the grid asset and determines whether the behavior deviates from expected operation. Its role is to produce early abnormality evidence, not final attack attribution.

The output of this layer summarizes the observed anomaly, the strength of the deviation, relevant communication context, and the features that contributed to the alert. This evidence is forwarded to the cloud reasoning layer, where it is combined with interpretation, model explanation features, and threat intelligence to support ranked cyber threat hypotheses.

\subsection{Cloud Threat Reasoning Layer}

The cloud layer transforms edge-generated abnormality evidence into structured cyber threat interpretation. Its role is to examine whether the alert is consistent with known or plausible attack behavior, enrich the evidence with model explanation features and CTI, and organize the result for attack-tree generation.

Because a single alert may contain indicators of multiple attack behaviors, the cloud reasoning layer does not treat classification as a final decision. Instead, it supports ranked threat hypotheses based on the available evidence, its quality, and the remaining uncertainty.

\subsection{Model Explanation Layer}

The model explanation layer extracts explicit reasoning from the cloud classifier's decisions by combining SHAP feature attributions with surrogate decision rules. Its primary role is to expose the exact behavioral features driving a classification, ensuring the threat hypothesis is grounded in interpretable model behavior rather than opaque scores. While the layer includes a secondary consistency check between SHAP and rule evidence to generate an evidence-quality signal, the main contribution to the unified evidence packet is the detailed feature-level and rule-based explanation. This allows the reasoning process to preserve uncertainty when evidence is weak or incomplete and to assign stronger support when the model explanation is clear and consistent. In this way, robust model explanations help reduce unsupported conclusions and strengthen evidence-grounded threat reasoning.

\subsection{Cyber Threat Intelligence Context Layer}

The CTI layer adds external context to the alert by linking observed behavior to known adversary techniques, attack patterns, and official mitigation guidance. In CyTReX, CTI retrieval is behavior-oriented: MITRE ATT\&CK for ICS provides candidate techniques, tactics, and associated mitigations, while CAPEC provides related attack-pattern context. 

This layer is accessed through an MCP-enabled interface \cite{mcp_architecture} and contributes provenance-aware CTI evidence to the unified evidence packet. This design keeps the reasoning process traceable and prevents unsupported CTI claims when the alert provides only behavior-level evidence.

\subsection{Unified Evidence Packet and Attack-Tree Reasoning}

The unified evidence packet serves as the structured input for LLM-based reasoning. It brings together edge evidence, cloud reasoning layer attack interpretation, model explanation features, CTI, evidence-quality signals, and provenance metadata.

Using this packet, the LLM generates an attack tree that includes the root suspicious event, ranked threat hypotheses, supporting evidence, uncertainty notes, and evidence gaps. The output is intended to support analyst interpretation and decision readiness, not automated response execution.

\begin{figure}[t]
    \centering
    \includegraphics[width=\columnwidth]{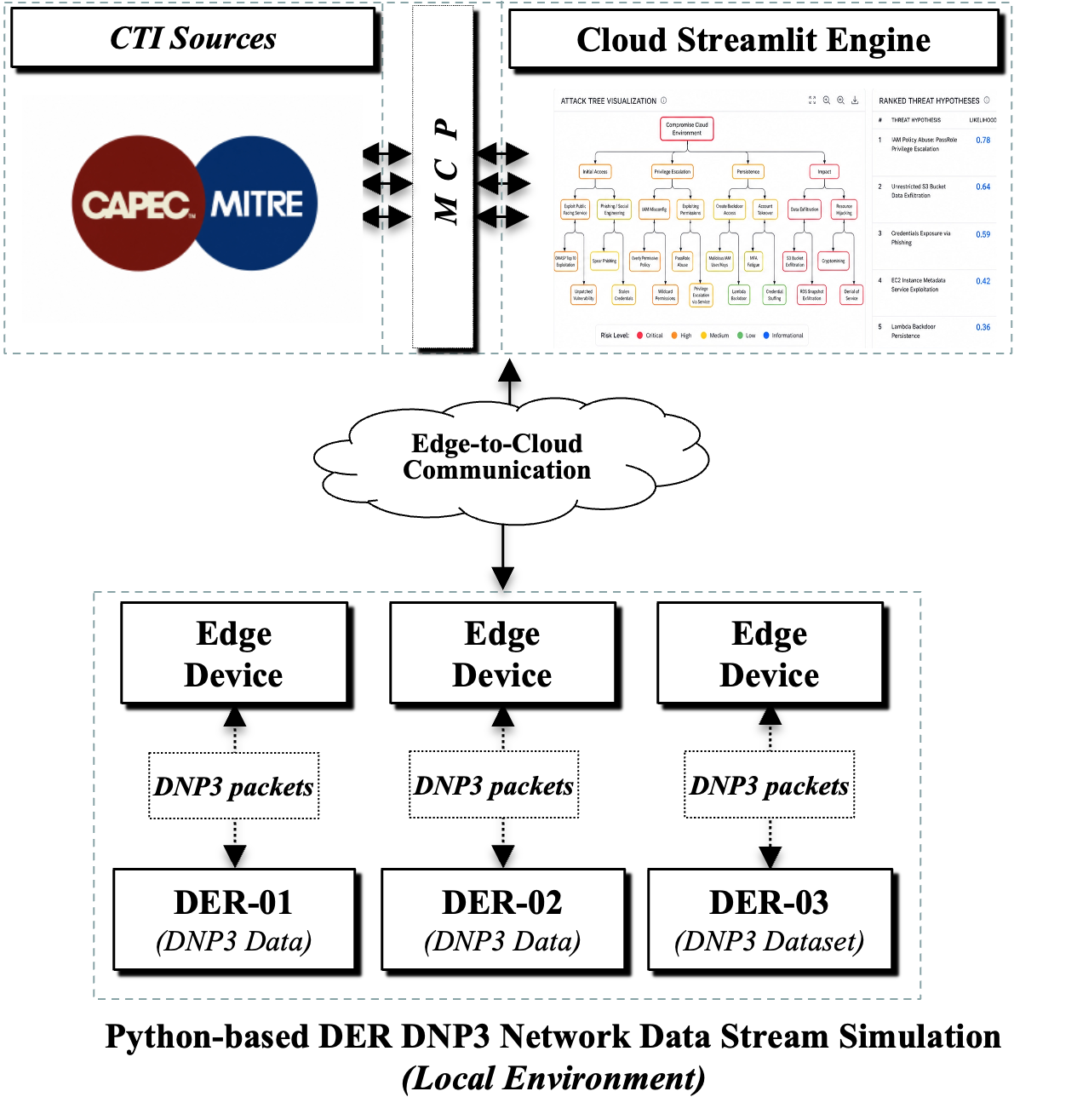}
    \vspace{-20pt}
    \caption{Experimental setup showing DER devices transmitting DNP3 packets to edge devices, with edge-to-cloud communication, MCP-enabled CTI retrieval, and cloud-based threat reasoning through the Streamlit engine.}
    \label{fig:experimental_setup}
    \vspace{-15pt}
\end{figure}

\section{Experimental Evaluation}

\subsection{Experimental Setup}

Fig.~\ref{fig:experimental_setup} illustrates the experimental setup, which emulates a DER network environment.  CyTReX is implemented in Python~3.13 on an Apple M4 system (16~GB of random-access memory (RAM), macOS~15.5) with a Streamlit-based interface running Ollama model locally. Three virtual DER devices simulate DNP3 traffic in real time to the edge detection layer, where a semi-supervised anomaly detector is used to identify abnormal communication behavior and aggregate alerts into JavaScript Object Notation (JSON) batches for cloud layer processing. The cloud reasoning layer uses a supervised classifier for attack-class interpretation, SHAP for feature-attribution evidence, surrogate rule evidence for explanation verification, and an MCP-enabled CTI interface for external context retrieval. The LLM layer uses \texttt{llama3.1:8b} through a local Ollama backend for all attack-tree generation tasks. The same model is used across all five evaluation configurations, with no configuration-specific LLM changes, so output differences are attributed to the active evidence components rather than to different LLM settings.

\subsection{Dataset and Case-Study Labels}

The evaluation uses a DNP3-based communication dataset as the case-study environment \cite{radoglou2022dnp3}. Table~\ref{tab:dataset_labels} summarizes the dataset labels and their meaning in a DNP3 communication context. The supervised cloud classifier is trained on six attack labels and normal traffic, while \texttt{DNP3\_ENUM} and \texttt{DNP3\_INFO} are excluded from training and used as held-out out-of-distribution (OOD) cases for reconnaissance-style behavior. A representative edge-device alert batch is used across all configurations so that each setting reasons over the same cyber-physical event. Although DNP3 provides the case-study protocol, CyTReX operates on normalized evidence objects rather than protocol-specific assumptions.

For model selection, the dataset was split into training and testing subsets (70:30) before evaluating the edge and cloud models reported in Table~\ref{tab:ml_results}. The edge anomaly detector was trained to learn normal communication behavior and tested on normal and attack traffic, while the cloud classifier was trained and tested using the closed-set labels in Table~\ref{tab:dataset_labels}. This separation ensures that Table~\ref{tab:ml_results} reports closed-set model performance, while the held-out OOD cases evaluate reasoning behavior outside the trained attack-label space.

\begin{table*}[t]
  \centering
  \caption{Dataset labels in the DNP3 case study.}
  \label{tab:dataset_labels}
  \scriptsize
  \setlength{\tabcolsep}{3pt}
  \renewcommand{\arraystretch}{1.10}
  \begin{tabular}{p{0.15\textwidth}p{0.16\textwidth}p{0.62\textwidth}}
    \hline
    Label & Use in this study & Meaning in DNP3 context \\
    \hline
    \texttt{NORMAL} & Benign baseline & Expected DNP3 communication behavior. \\
    \texttt{REPLAY} & Trained attack & Replay of DNP3 packets previously captured from a legitimate DNP3 master or outstation. \\
    \texttt{COLD\_RESTART} & Trained attack & A malicious master-like entity sends the DNP3 cold-restart function code, causing the target outstation to initiate a complete restart. \\
    \texttt{WARM\_RESTART} & Trained attack & A malicious master-like entity sends the DNP3 warm-restart function code, causing a partial restart or reinitialization of the DNP3 service. \\
    \texttt{STOP\_APP} & Trained attack & A stop-application command associated with DNP3 Function Code 18 asks the slave/outstation to stop a function so it cannot continue receiving messages from the master. \\
    \texttt{DISABLE\_UNSOLICITED} & Trained attack & A command associated with DNP3 Function Code 21 disables unsolicited messages from the target outstation. \\
    \texttt{INIT\_DATA} & Trained attack & An initialize-data command associated with DNP3 Function Code 15 requests unauthorized reinitialization of device data or configuration values. \\
    \texttt{DNP3\_ENUM} & Held-out OOD test & Reconnaissance activity intended to identify which DNP3 services and function codes are supported by the target system. \\
    \texttt{DNP3\_INFO} & Held-out OOD test & Reconnaissance activity that collects DNP3 diagnostic or usage information from the target environment. \\
    \hline
  \end{tabular}
  \vspace{-15pt}
\end{table*}

\begin{table}[t]
  \centering
  \caption{MCP-enabled CTI tools used in the evaluation.}
  \label{tab:mcp-tools}
  \scriptsize
  \setlength{\tabcolsep}{2.5pt}
  \renewcommand{\arraystretch}{1.08}
  \begin{tabular}{p{0.41\linewidth}p{0.49\linewidth}}
    \hline
    \textbf{Tool} & \textbf{Purpose} \\
    \hline
    \texttt{extract\_evidence\_terms}    & Extracts CTI query terms from edge, cloud, SHAP, rule, and flow evidence. \\
    \texttt{query\_mitre\_attack\_ics}   & Retrieves MITRE ATT\&CK for ICS techniques, tactics, STIX relationships, and official mitigations. \\
    \texttt{query\_capec\_patterns}      & Retrieves CAPEC attack-pattern context from the CAPEC Extensible Markup Language (XML) corpus. \\
    \texttt{retrieve\_full\_cti\_context} & Orchestrates evidence-term extraction, MITRE retrieval, CAPEC retrieval, normalization, and provenance recording. \\
    \hline
  \end{tabular}
\end{table}

\subsection{Model Explanation and Feature Attribution}

The Rule-SHAP consistency formulation is adapted from prior work \cite{excyder2026}, while the surrogate rule-quality score is introduced here as an additional verification signal. CyTReX extracts the unique feature names from the surrogate rules, denoted as $F_{\mathrm{r}}$, and uses their count to set the SHAP evidence budget:
\begin{equation}
k = \max(1, |F_{\mathrm{r}}|),
\end{equation}
where $k$ is the number of top-ranked SHAP features retained for comparison and $F_{\mathrm{s}}^{(k)}$ denotes the corresponding top-$k$ SHAP feature set. The Rule-SHAP consistency score combines feature overlap $O$, direction agreement $D$, and rule coverage $C$:
\begin{equation}
S_{\mathrm{rs}} = \alpha O + \beta D + \gamma C,
\end{equation}
where $\alpha=0.45$, $\beta=0.10$, and $\gamma=0.45$. The larger weights on feature overlap and rule coverage place the score mainly on feature-level agreement: overlap captures whether SHAP and surrogate rules identify similar influential features, while coverage captures whether rule-derived features are represented in the SHAP evidence budget. Direction agreement receives a lower weight because SHAP contribution direction and surrogate branch direction may differ even when both identify the same behaviorally relevant feature.

CyTReX also computes a surrogate rule-quality score:
\begin{equation}
Q_{\mathrm{r}} = \eta F_{\mathrm{sur}} + (1-\eta)R_{\mathrm{d}},
\end{equation}
where $F_{\mathrm{sur}}$ is surrogate fidelity and $R_{\mathrm{d}}$ is rule density relative to the SHAP budget. In this study, $\eta=0.65$, giving fidelity a moderate majority weight over rule density because rule explanations are most useful when the surrogate closely approximates the cloud classifier. Rule density retains a secondary weight of $0.35$ to reward explanation availability without treating more rules as inherently better explanations. Both $S_{\mathrm{rs}}$ and $Q_{\mathrm{r}}$ are normalized scores in the range $[0,1]$, where values closer to $1$ indicate stronger explanation support and values closer to $0$ indicate weaker support. Rule-only features that do not overlap with SHAP are retained because they may expose additional surrogate decision patterns.

The verified confidence is computed as:
\begin{equation}
\mathrm{Conf}*{\mathrm{v}} = \mathrm{Conf}*{\mathrm{m}} \times (0.5 + 0.5V),
\end{equation}
where $\mathrm{Conf}*{\mathrm{m}}$ is the original model confidence and $V=\max(S*{\mathrm{rs}}, 0.5Q_{\mathrm{r}})$ is the combined verification signal. The weighting values are design parameters, and future work will assess how alternative schemes affect hypothesis ranking, evidence-quality labels, and reasoning consistency.

\subsection{MCP-Enabled CTI Retrieval}

The MCP layer adds structured CTI context to the unified evidence packet through a FastMCP-based client-server module. Its scope is limited to CTI enrichment and provenance-aware retrieval, and it does not perform anomaly detection, attack classification, SHAP computation, surrogate-rule extraction, attack-tree generation, or automated mitigation. In the evaluation, MCP enrichment is enabled only in the edge-plus-CTI and complete-pipeline configurations. Rather than allowing the LLM to perform unconstrained threat-intelligence lookup, the MCP client converts alert evidence into structured CTI requests by extracting discriminative behavioral terms from edge alerts, cloud classifications, SHAP-ranked features, surrogate decision-rule features, and flow metadata, while filtering generic domain vocabulary such as DNP3, ICS, and supervisory control and data acquisition (SCADA).

The server retrieves behavior-level CTI from MITRE ATT\&CK for ICS and CAPEC. MITRE ATT\&CK for ICS provides adversary techniques, tactics, STIX relationships, and official mitigation guidance, while CAPEC provides complementary attack-pattern context. The server queries MITRE ATT\&CK for ICS through the TAXII 2.1 endpoint, matches active STIX techniques against evidence-derived terms, and resolves official mitigations through the STIX \texttt{mitigates} relationship. CAPEC patterns are retrieved from the CAPEC XML corpus and scored using the same evidence-derived terms. The retrieved techniques, attack patterns, mitigations, source references, retrieval status, and provenance metadata are packaged into a unified CTI context for LLM-based reasoning. In the local evaluation, live retrieval is used with caching disabled and in-process transport as the default, while standard input/output (\texttt{stdio}), Hypertext Transfer Protocol (HTTP), and Server-Sent Events (SSE) transport modes are also supported.

\subsection{Evaluation Configurations}

The evaluation uses five configurations grouped into two tracks. The edge-only track comprises configurations C1 and C2 and evaluates how much reasoning can be supported from edge evidence alone and how CTI changes that edge-only reasoning. The full-pipeline track comprises configurations C3, C4, and C5 and evaluates the added value of cloud reasoning layer attack interpretation, model explanation features, CTI enrichment, and the complete evidence-grounded reasoning workflow. Table~\ref{tab:evaluation_configs} summarizes the configurations.

\begin{table}[t]
  \centering
  \caption{Five-configuration evaluation design.}
  \label{tab:evaluation_configs}
  \scriptsize
  \setlength{\tabcolsep}{2.5pt}
  \renewcommand{\arraystretch}{1.12}
  \begin{tabular}{p{0.12\linewidth}p{0.43\linewidth}p{0.35\linewidth}}
    \hline
    Config. & Configuration & Purpose \\
    \hline
    C1 & Edge-only: Edge evidence + LLM & Tests whether edge anomaly evidence alone supports attack-tree generation. \\
    C2 & Edge-only: Edge evidence + CTI + LLM & Tests whether CTI enriches edge-only reasoning without cloud reasoning layer attack interpretation. \\
    C3 & Full pipeline: Edge evidence + cloud reasoning layer attack interpretation + LLM & Tests the effect of cloud reasoning layer attack interpretation. \\
    C4 & Full pipeline: Edge evidence + cloud reasoning layer attack interpretation + Rule-SHAP + LLM & Tests whether model explanation features improves evidence support. \\
    C5 & Complete pipeline: Edge evidence + cloud + Rule-SHAP + CTI + LLM & Tests the complete evidence-grounded reasoning workflow. \\
    \hline
  \end{tabular}
  \vspace{-15pt}
\end{table}

For the ranked outputs, H1 denotes the highest-ranked hypothesis, while H2 and H3 denote the second- and third-ranked hypotheses, respectively.

\begin{table*}[t]
  \centering
  \caption{Overview of generated highest-ranked (H1) hypotheses and latency across trained DNP3 attack labels.}
  \label{tab:config_perf_summary}
  \scriptsize
  \setlength{\tabcolsep}{1.5pt}
  \renewcommand{\arraystretch}{1.12}
  \begin{tabular}{p{0.16\textwidth}p{0.07\textwidth}p{0.15\textwidth}p{0.15\textwidth}p{0.15\textwidth}p{0.15\textwidth}p{0.15\textwidth}}
    \hline
    Attack label & Result & C1 & C2 & C3 & C4 & C5 \\
    \hline
    \multirow{2}{*}{\texttt{COLD\_RESTART}} & H1 & DNP3 replay & Rogue master & Cold restart & Cold restart & Denial of service \\
     & Latency & 152.06 s & 180.74 s & 138.85 s & 117.95 s & 160.33 s \\
    \hline
    \multirow{2}{*}{\texttt{DISABLE\_UNSOLICITED}} & H1 & Command replay & Distributed denial-of-service (DDoS) attack & Disable unsolicited & Cold restart & DNP3 replay \\
     & Latency & 135.82 s & 182.78 s & 122.78 s & 151.20 s & 134.18 s \\
    \hline
    \multirow{2}{*}{\texttt{INIT\_DATA}} & H1 & Rogue master & Denial of service & Rogue master & Cold restart & DNP3 replay \\
     & Latency & 150.28 s & 157.64 s & 142.43 s & 176.39 s & 133.37 s \\
    \hline
    \multirow{2}{*}{\texttt{REPLAY}} & H1 & DNP3 replay & DNP3 replay & DNP3 replay & DNP3 replay & DNP3 replay \\
     & Latency & 158.71 s & 172.93 s & 132.41 s & 145.39 s & 172.38 s \\
    \hline
    \multirow{2}{*}{\texttt{STOP\_APP}} & H1 & Injection & DNP3 replay & Device restart & Cold restart & Command injection \\
     & Latency & 295.72 s & 181.80 s & 126.10 s & 186.58 s & 142.61 s \\
    \hline
    \multirow{2}{*}{\texttt{WARM\_RESTART}} & H1 & DNP3 manipulation & Denial of service & Warm restart & Cold restart & Denial of service \\
     & Latency & 150.15 s & 373.03 s & 184.95 s & 167.56 s & 151.29 s \\
    \hline
  \end{tabular}
\end{table*}

\subsection{Evaluation Criteria}

The evaluation considers two aspects: ML detection performance and threat-reasoning behavior. Detection performance is measured using accuracy, precision, recall, the F1-score (the harmonic mean of precision and recall), confusion matrix, and prediction latency to verify the reliability of the edge and cloud model outputs used as evidence. Threat-reasoning behavior is evaluated using the primary H1 hypothesis, runtime latency, evidence grounding, and uncertainty handling across the five configurations. The discussion focuses on whether each configuration preserves the dataset label or a close semantic interpretation, whether the generated hypothesis aligns with the DNP3 behavior described in Table~\ref{tab:dataset_labels}, and whether additional components such as cloud reasoning layer attack interpretation, Rule-SHAP verification, and CTI enrichment improve the traceability of the reasoning output. For held-out \texttt{DNP3\_ENUM} and \texttt{DNP3\_INFO} cases, the evaluation examines whether the framework treats unseen reconnaissance-like behavior as uncertain rather than forcing unsupported closed-set conclusions.

% ----------------------------------------------------------------
% SECTION VI — RESULTS AND DISCUSSION
% ----------------------------------------------------------------

\section{Results and Discussion}

\subsection{ML Performance and Model Selection}

Table~\ref{tab:ml_results} reports the model-selection results for the edge and cloud layers using accuracy, precision, recall, and F1-score, abbreviated as Acc., Prec., Rec., and F1, respectively. Accuracy measures the overall proportion of correct predictions, precision measures the correctness of predicted attack or anomaly cases, recall measures how many true attack or anomaly cases are detected, and F1-score summarizes the balance between precision and recall, with higher values indicating stronger performance. Based on these results, the Isolation Forest and nearest-neighbor ensemble was selected for edge-layer abnormality detection, while Light Gradient Boosting Machine (LightGBM) was selected for cloud-layer supervised attack-behavior interpretation.

\begin{table}[t]
  \centering
  \caption{ML model selection results for edge and cloud reasoning layers.}
  \label{tab:ml_results}
  \scriptsize
  \setlength{\tabcolsep}{2.5pt}
  \renewcommand{\arraystretch}{1.08}
  \begin{tabular}{p{0.22\linewidth}p{0.32\linewidth}cccc}
    \hline
    Layer & Model & Acc. & Prec. & Rec. & F1 \\
    \hline
    Edge & Isolation Forest+Nearest Neighbor & 98.43 & 99.04 & 99.37 & 99.21 \\
    Edge & Isolation Forest & 97.93 & 98.98 & 98.85 & 98.91 \\
    Edge & Extended Isolation Forest & 57.52 & 98.50 & 56.26 & 71.61 \\
    \hline
    Cloud & LightGBM & 99.58 & 99.58 & 99.58 & 99.58 \\
    Cloud & Extreme Gradient Boosting (XGBoost) & 99.08 & 99.10 & 99.08 & 99.08 \\
    Cloud & Random Forest & 85.79 & 88.31 & 85.80 & 85.32 \\
    \hline
  \end{tabular}
  \vspace{-15pt}
\end{table}

\subsection{Attack-Label Reasoning Results and Latency}

Table~\ref{tab:config_perf_summary} reports the primary H1 hypothesis and runtime latency for each trained DNP3 attack label across C1--C5. These outputs are interpreted as threat-reasoning results rather than direct classifier predictions using three levels of interpretation. An exact label match occurs when H1 preserves the dataset label or a direct variant, such as \texttt{REPLAY} being reported as DNP3 replay. A semantic match occurs when H1 does not repeat the dataset label but captures the same attacker role, command action, or operational effect, such as restart behavior being mapped to denial of service or \texttt{STOP\_APP} being mapped to command injection. A mismatch occurs when H1 shifts to a weakly related behavior, such as C1 mapping \texttt{COLD\_RESTART} to DNP3 replay, indicating that the available evidence is insufficient for precise attack attribution.

The results show that edge-only configurations produce broader hypotheses because they lack supervised attack-class context, while CTI can widen the hypothesis space by mapping protocol-specific DNP3 labels to broader behavior-level concepts. C3 provides the strongest label anchoring by preserving \texttt{COLD\_RESTART}, \texttt{DISABLE\_UNSOLICITED}, \texttt{REPLAY}, and \texttt{WARM\_RESTART} as H1. \texttt{REPLAY} appears as H1 under all five configurations, showing strong behavioral separability, whereas \texttt{INIT\_DATA} is not preserved as H1, suggesting overlap with broader command or access-related behavior. C5 provides richer semantic interpretation for \texttt{STOP\_APP} and restart-related cases, but it may also shift from protocol-specific labels toward broader threat interpretations when CTI and explanation evidence are combined. These H1 mismatches show where the reasoning pipeline needs stronger packet-level, temporal, or open-set support. C5 appears weaker than C3 under exact label matching because C3 is mainly guided by the supervised cloud classifier, while C5 combines classifier evidence with Rule-SHAP and CTI context. This can shift protocol-specific labels toward broader threat effects, such as restart cases mapping to denial of service or \texttt{STOP\_APP} mapping to command injection. Therefore, C5 trades some label anchoring for richer evidence-grounded interpretation, mitigation context, and visibility into possible co-occurring behaviors.

Latency across C1--C5 is not strictly monotonic because runtime reflects the combined cost of LLM generation, CTI retrieval, evidence-packet size, and attack-tree complexity. C3 has the lowest average latency, while C2 has the highest mainly because the \texttt{WARM\_RESTART} case produced broader CTI matches that increased evidence-packet size and generation time. C5 remains relatively stable despite using the complete pipeline, showing that latency is better interpreted as an end-to-end reasoning cost than as a direct function of the number of active components. This also clarifies the role of the LLM in CyTReX. The LLM is not evaluated as a standalone detector or unconstrained CTI search engine, but as a controlled reasoning layer that converts each configuration's evidence packet into ranked hypotheses and attack-tree branches. Its contribution is therefore assessed through H1 correctness, semantic alignment, evidence grounding, uncertainty handling, and end-to-end latency.

\begin{table}[!t]
\centering
\caption{OOD reasoning results for held-out DNP3 labels.}
\label{tab:ood_reasoning}
\scriptsize
\renewcommand{\arraystretch}{1.05}
\begin{tabular}{llcc}
\toprule
\textbf{OOD label} & \textbf{Result} & \textbf{C2} & \textbf{C5} \\
\midrule
\texttt{DNP3\_ENUM}
& H1 & Parameter manipulation & DNP3 replay \\
& H2 & Data collection & Denial of service \\
& H3 & Process disruption & Rogue master injection \\
& Latency & 153.00 s & 136.15 s \\
\midrule
\texttt{DNP3\_INFO}
& H1 & Parameter manipulation & DNP3 replay \\
& H2 & Initial access attempt & DNP3 spoofing \\
& H3 & Lateral movement attempt & DNP3 command control \\
& Latency & 202.84 s & 135.85 s \\
\bottomrule
\vspace{-20pt}
\end{tabular}
\end{table}

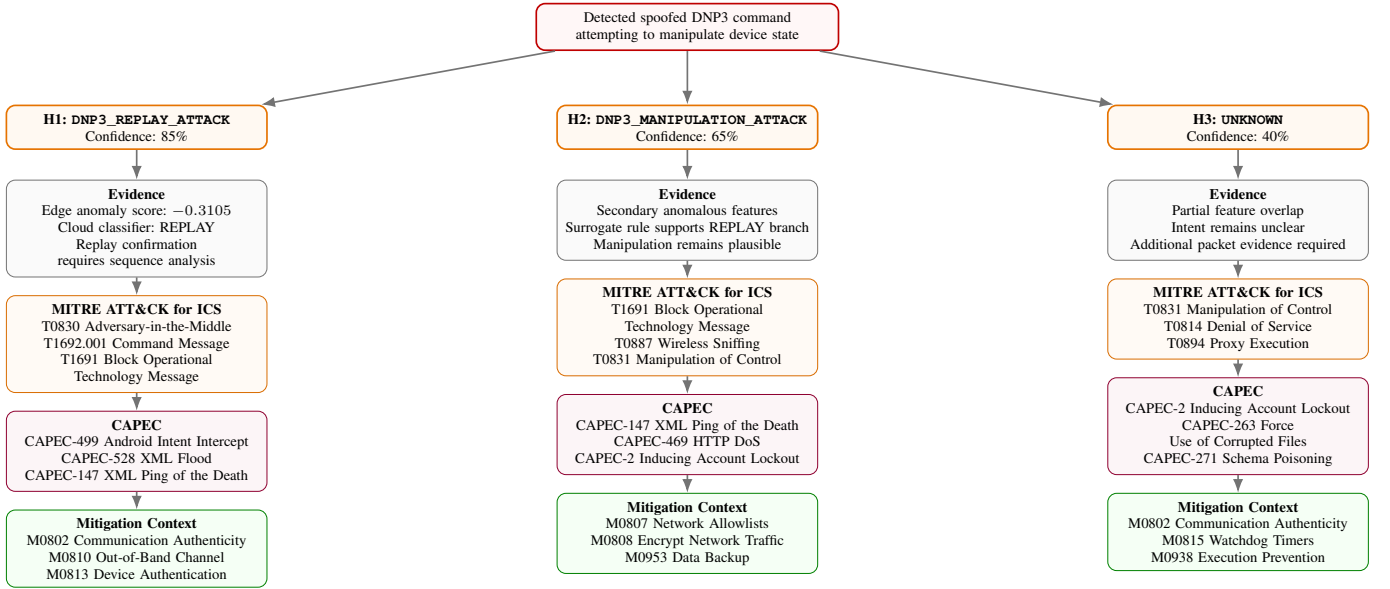
\begin{figure*}[t]
\centering
\resizebox{\textwidth}{!}{%
\begin{tikzpicture}[
    font=\scriptsize,
    box/.style={
        draw=black!55,
        rounded corners=1.5mm,
        align=center,
        fill=black!2,
        inner sep=4pt
    },
    root/.style={
        box,
        draw=red!75!black,
        thick,
        fill=red!3,
        text width=4.8cm
    },
    hyp/.style={
        box,
        draw=orange!90!black,
        thick,
        fill=orange!5,
        text width=4.1cm
    },
    ev/.style={
        box,
        draw=black!55,
        fill=gray!4,
        text width=4.1cm
    },
    mitre/.style={
        box,
        draw=orange!85!black,
        fill=orange!4,
        text width=4.1cm
    },
    capec/.style={
        box,
        draw=purple!75!black,
        fill=purple!4,
        text width=4.1cm
    },
    mit/.style={
        box,
        draw=green!50!black,
        fill=green!4,
        text width=4.1cm
    },
    arrow/.style={-Latex, thick, draw=black!55}
]

% Root event
\node[root] (root) {
Detected spoofed DNP3 command attempting to manipulate device state
};

% Hypotheses
\node[hyp, below left=9mm and 45mm of root] (h1) {
\textbf{H1: \texttt{DNP3\_REPLAY\_ATTACK}}\\
Confidence: 85\%
};

\node[hyp, below=9mm of root] (h2) {
\textbf{H2: \texttt{DNP3\_MANIPULATION\_ATTACK}}\\
Confidence: 65\%
};

\node[hyp, below right=9mm and 45mm of root] (h3) {
\textbf{H3: \texttt{UNKNOWN}}\\
Confidence: 40\%
};

\draw[arrow] (root) -- (h1);
\draw[arrow] (root) -- (h2);
\draw[arrow] (root) -- (h3);

% H1 column
\node[ev, below=5mm of h1] (h1e) {
\textbf{Evidence}\\
Edge anomaly score: $-0.3105$\\
Cloud classifier: REPLAY\\
Replay confirmation requires sequence analysis
};

\node[mitre, below=3mm of h1e] (h1m) {
\textbf{MITRE ATT\&CK for ICS}\\
T0830 Adversary-in-the-Middle\\
T1692.001 Command Message\\
T1691 Block Operational Technology Message
};

\node[capec, below=3mm of h1m] (h1c) {
\textbf{CAPEC}\\
CAPEC-499 Android Intent Intercept\\
CAPEC-528 XML Flood\\
CAPEC-147 XML Ping of the Death
};

\node[mit, below=3mm of h1c] (h1r) {
\textbf{Mitigation Context}\\
M0802 Communication Authenticity\\
M0810 Out-of-Band Channel\\
M0813 Device Authentication
};
\draw[arrow] (h1) -- (h1e);
\draw[arrow] (h1e) -- (h1m);
\draw[arrow] (h1m) -- (h1c);
\draw[arrow] (h1c) -- (h1r);

% H2 column
\node[ev, below=5mm of h2] (h2e) {
\textbf{Evidence}\\
Secondary anomalous features\\
Surrogate rule supports REPLAY branch\\
Manipulation remains plausible
};

\node[mitre, below=3mm of h2e] (h2m) {
\textbf{MITRE ATT\&CK for ICS}\\
T1691 Block Operational Technology Message\\
T0887 Wireless Sniffing\\
T0831 Manipulation of Control
};

\node[capec, below=3mm of h2m] (h2c) {
\textbf{CAPEC}\\
CAPEC-147 XML Ping of the Death\\
CAPEC-469 HTTP DoS\\
CAPEC-2 Inducing Account Lockout
};

\node[mit, below=3mm of h2c] (h2r) {
\textbf{Mitigation Context}\\
M0807 Network Allowlists\\
M0808 Encrypt Network Traffic\\
M0953 Data Backup
};
\draw[arrow] (h2) -- (h2e);
\draw[arrow] (h2e) -- (h2m);
\draw[arrow] (h2m) -- (h2c);
\draw[arrow] (h2c) -- (h2r);

% H3 column
\node[ev, below=5mm of h3] (h3e) {
\textbf{Evidence}\\
Partial feature overlap\\
Intent remains unclear\\
Additional packet evidence required
};

\node[mitre, below=3mm of h3e] (h3m) {
\textbf{MITRE ATT\&CK for ICS}\\
T0831 Manipulation of Control\\
T0814 Denial of Service\\
T0894 Proxy Execution
};

\node[capec, below=3mm of h3m] (h3c) {
\textbf{CAPEC}\\
CAPEC-2 Inducing Account Lockout\\
CAPEC-263 Force Use of Corrupted Files\\
CAPEC-271 Schema Poisoning
};

\node[mit, below=3mm of h3c] (h3r) {
\textbf{Mitigation Context}\\
M0802 Communication Authenticity\\
M0815 Watchdog Timers\\
M0938 Execution Prevention
};
\draw[arrow] (h3) -- (h3e);
\draw[arrow] (h3e) -- (h3m);
\draw[arrow] (h3m) -- (h3c);
\draw[arrow] (h3c) -- (h3r);

\end{tikzpicture}
}
\vspace{-20pt}
\caption{Compact evidence-grounded attack-tree representation for the \texttt{REPLAY} case, showing the root event, ranked threat hypotheses, supporting evidence, MITRE ATT\&CK for ICS techniques, CAPEC patterns, and mitigation context.}
\label{fig:c5_replay_attack_tree}
%\vspace{-15pt}
\end{figure*}

\subsection{Replay Attack-Tree Analysis}

Fig.~\ref{fig:c5_replay_attack_tree} shows the attack tree generated for the \texttt{REPLAY} case using the C5 complete-pipeline configuration. This case is representative because \texttt{REPLAY} is a well-defined DNP3 threat, appears as H1 across all configurations in Table~\ref{tab:config_perf_summary}, and still requires evidence-grounded interpretation beyond label matching. The generated tree assigns \texttt{DNP3\_REPLAY\_ATTACK} as the primary hypothesis and \texttt{DNP3\_MANIPULATION\_ATTACK} as a secondary hypothesis, since a replayed command may manipulate device state, disrupt control flow, or trigger an unauthorized response. The \texttt{UNKNOWN} hypothesis preserves uncertainty where packet-sequence evidence is insufficient to fully confirm replay.

The retrieved MITRE ATT\&CK for ICS context frames the alert around replay-related behaviors, including command manipulation, unauthorized control activity, and operational disruption. The CTI and mitigation entries therefore provide behavior-level support and investigation guidance, not direct proof that every retrieved technique or pattern occurred. Mitigation overlap across H1, H2, and H3, including communication authenticity, authentication, network allowlisting, and out-of-band validation, can support action prioritization by highlighting defenses that remain relevant across multiple plausible hypotheses. Although Fig.~\ref{fig:c5_replay_attack_tree} presents the detailed replay example, Table~\ref{tab:config_perf_summary} also shows similar reasoning for other labels, where \texttt{STOP\_APP} maps to command injection and restart-related cases map to denial-of-service-oriented behavior, reflecting their operational effects.

\subsection{Out-of-Distribution Behavior}

The held-out \texttt{DNP3\_ENUM} and \texttt{DNP3\_INFO} cases are treated as OOD tests. As shown in Table~\ref{tab:ood_reasoning}, CyTReX generates security-relevant hypotheses for both unseen reconnaissance-style behaviors. In C2, edge anomaly evidence and CTI context produce broad hypotheses such as parameter manipulation, data collection, initial access, and lateral movement. In C5, cloud interpretation, SHAP attribution, and surrogate rule evidence enrich the output, but the closed-set classifier can also pull unseen behavior toward known classes such as replay. This confirms that CyTReX provides transparent reasoning for OOD cases while highlighting the need for explicit open-set detection to avoid forcing unseen behaviors into known attack categories.

% ----------------------------------------------------------------
% SECTION VII — CONCLUSION
% Varied from abstract to avoid verbatim repetition
% ----------------------------------------------------------------
\section{Conclusion}

This paper presented CyTReX, an evidence-grounded cybersecurity threat reasoning framework for DER networks that addresses a key limitation of single-label anomaly detection: a detection label reports the selected class but does not expose alternative threat interpretations or the evidence needed to distinguish them. CyTReX moves beyond this limitation by generating ranked, traceable threat hypotheses from integrated detection evidence, XAI-based network-feature explanations, MCP-enabled CTI enrichment, and LLM-based attack-tree reasoning, providing SOC analysts with evidence context for informed triage and response. The five-configuration evaluation shows that edge-only reasoning detects abnormal behavior but remains limited for attack attribution, while cloud reasoning, model explanation, and CTI enrichment improve evidence traceability and analytical grounding. The OOD evaluation further shows that CyTReX can generate security-relevant hypotheses for unseen reconnaissance-like behavior, while also highlighting the need for explicit open-set handling. Future work will evaluate CyTReX on larger datasets and broader DER communication scenarios, and extend it toward human-in-the-loop response support where evidence-grounded reasoning can inform analyst-approved mitigation actions.

\section*{Acknowledgement}
This work is funded in part by the U.S. Department of Energy (DOE), Office of Cybersecurity, Energy Security, and Emergency Response (CESER), under Award Number DE-CR0000049.

\balance
\bibliographystyle{IEEEtran}
\bibliography{references_corrected}

\end{document}